\documentstyle[12pt,fleqn,leqno]{article}

\begin{document}

\title{Boltzmann Collision Kernels and Velocity Saturation
in Semiconductors}
\author{Christoph Dalitz\\
Universit\"at Bielefeld, Theoretische Physik\\
Universit\"atsstr.~25, D-33615 Bielefeld}
\date{}
\maketitle
\begin{abstract}
For different models of the electron-phonon interaction,
the asymptotic behaviour of the moments of the stationary 
homogeneous solution of the linear Boltzmann equation is determined in the 
limit of a high external field. For Hilbert-Schmidt kernels of a finite
rank, a result recently
proven for kernels of rank one is found generally valid; as a consequence
velocity saturation is excluded for these collision models. For a class of
singular collision kernels in contrast, velocity saturation is generally
obtained.
\end{abstract}

%
%
\newcounter{saveeqn}
\newcommand{\subeqn}{\refstepcounter{equation}%
 \setcounter{saveeqn}{\value{equation}}%
 \setcounter{equation}{0}%
   \renewcommand{\theequation}
   {\mbox{\arabic{saveeqn}\alph{equation}}}}
\newcommand{\reseteqn}{\setcounter{equation}{\value{saveeqn}}%
 \renewcommand{\theequation}{\mbox{\arabic{equation}}}}
\newenvironment{subeqnarray}{%
\subeqn\begin{eqnarray}}
{\end{eqnarray}\reseteqn}
%
%
%
\newcommand{\R}{\mbox{{\rm I\hspace{-0.6mm}R}}}

\section{Introduction}
It is a well known fact that the drift velocity of electrons in
a semiconductor approaches a finite value, if the applied field
becomes very large. Already {\sc Shockley}, whose experiments
with {\sc Ryder} \cite{Ryder} first revealed saturation of
the electron drift velocity in germanium, argued in an intuitive
manner that optical phonon scattering should be responsible for
this effect \cite{Shockley}. Meanwhile, changeover from acoustic
to optic mode phonon emission in high electric fields has been
examined experimentally in some detail \cite{Hawker}. \par
For a mathematical explanation of velocity saturation, it is 
necessary to determine the high field behaviour of the first moment
with respect to the momentum $p$ of the solution $f(p)$ of the 
stationary homogeneous Boltzmann equation \cite{Markowich}
\begin{equation}
E\frac{\partial}{\partial p}f(p) = {\cal Q}f(p)
\end{equation}
For several specific models of the electron-phonon collision
operator ${\cal Q}$, this behaviour has been determined by different 
authors \cite{Reik}, \cite{Gervois}, \cite{Dalitz_moments}. 
Particularly {\sc Reik} and {\sc Risken} have shown that in the
presence of both optic and acoustic mode deformation potential
scattering the electron drift velocity saturates \cite{Reik}.
Nevertheless the interesting question remains as to which general
classes of collision models actually do or do not lead to
velocity saturation. \par
In this article we answer this question for two classes of collision
models within the framework of a parabolic band approximation 
and the assumption of a nondegenerate electron gas,
which means that the electron concentration be not too high 
\cite{Fuchs}. The main results are given in form of theorems at
the end of section three and five.\par
In section three and four, we examine regular kernels: if their
rank is finite, the high field behaviour of the moments is given by
a law which the author has proven recently for ``relaxation time
approximations'' (kernels of rank one) \cite{Dalitz_moments}. As a 
consequence, velocity saturation is excluded for these kernels. This 
result seems to remain valid for kernels of infinite rank as well, as 
we point out in section three. \par
In section five we shall see that the situation changes considerably,
if the quantum nature of the lattice vibrations is taken into account.
Under the assumption of a single phonon frequency (``{\sc Einstein}'s
model'', which is a good approximation for the optical phonon mode
\cite{Becker}) and some isotropy hypothesis upon the scattering
probability, it will turn out that the electron drift velocity
saturates, regardless of the detailed functional form of the
scattering probability.

\section{The collision operator}
If the electron concentration is low enough, so that interactions among
the electrons themselves (via collisions or {\sc Pauli}'s exclusion principle)
 may be neglected, the electron phonon collision
operator ${\cal Q}$ in (1) is linear and has the general form 
\cite{Markowich} (throughout this article we set the effective mass
of the electrons equal unity, so that momentum $p$ and velocity are
the same) 
\begin{equation}
{\cal Q}f(p) = M(p)\int dp'\,K(p',p)\,f(p') - f(p)\int dp'\,M(p')\,K(p,p')
\end{equation}
where $M(p):=(2\pi\theta)^{-1/2}e^{-p^2\!/2\theta}$ is the Maxwellian 
at temperature $\theta$ of the semiconductor lattice and the {\it collision
kernel} $K(p,p')$ is a symmetric and positive distribution. The second integral
in (2), which gives the total scattering rate, is usually called {\it collision
frequency} $\nu$
\begin{equation}
\nu(p):=\int dp'\,M(p')\,K(p,p')
\end{equation}
The collision kernel $K(p,p')$ strongly
depends on the model for the electron-lattice interaction. We can roughly 
divide these models into two classes: \par
The first class are classical models, which allow an arbitrary
amount of energy exchange between electrons and lattice; these models lead
to continuous collision kernels or (under the not very restrictive assumption
that $K$ be squareintegrable with weight $M$) {\it Hilbert-Schmidt kernels}. 
Nearly all existence and uniqueness results for the linear Boltzmann equation 
refer to collision kernels belonging to this class.\par
The second class are semiclassical models, which take into account the 
quantum nature of the lattice vibrations. These models lead to 
{\it singular kernels}, where energy conservation is represented by Dirac 
$\delta$-distributions in the collision kernel.

\section{Hilbert-Schmidt kernels of finite rank}
Let us begin our examinations with collision kernels that are
measurable functions with a finite norm
\begin{equation}
\| K\|^2_M := \int\!dp\int\!dp'\,M(p)\,M(p')\,
K^2(p,p') < \infty
\end{equation}
Since $K$ is symmetric, the kernel $K'(p,p')=\sqrt{M(p)}K(p,p')\sqrt{M(p')}$
is symmetric and squareintegrable. Hence it follows from standard Hilbert-Schmidt
theory \cite{Fenyo} that $K'$ has an eigenfunction representation which
converges in $L^2(\R^6)$. If we write this representation in 
terms of $K$, we find
\begin{equation}
K(p,p') = \sum_{n=0}^\infty \lambda_n\,
\varphi_n(p)\,\varphi_n(p')
\end{equation}
where the right hand side converges with respect to the norm (4). The $n$-th
eigenfunction $\varphi_n$ is a solution of the eigenvalue equation with 
{\it eigenvalue} $\lambda_n$
\begin{equation}
\int\!dp'\,M(p')\,K(p,p')\,\varphi_n(p') = 
\lambda_n\,\varphi_n(p)
\end{equation}   
The eigenfunctions $\{\varphi_n\}$ form an orthonormal system with respect to 
the $L^2$-scalar\-product with weight $M$:
\begin{equation}
\int\!dp\,\varphi_n(p)\,\varphi_m(p)\,M(p) = \delta_{nm}
\end{equation}
Now let us assume that the eigenvalue equation (6) has a solution only
for a finite number of eigenvalues $\{ \lambda_n \}_{n=1}^{n_{max}}$, which
means that the kernel $K(p,p')$ is of {\it finite rank}. In this case the 
eigenfunction representation (5) takes the form
\begin{equation}
K(p,p') = \sum_{n=1}^{n_{max}} \lambda_n\,\varphi_n(p)\,\varphi_n(p')
\end{equation}
For the special case $n_{max}=1$, (8) is identical with the relaxation
time approximation discussed in \cite{Dalitz_moments}. We shall see 
that the stationary homogeneous Boltzmann equation can be readily integrated 
for any finite $n_{max}$ and the resulting expression for the moments 
allows an asymptotic evaluation for large fields $E$.

\subsection{Solution of the Boltzmann equation}
With the collision kernel (8), the homogeneous stationary Boltzmann equation
takes the form
\begin{equation}
\Big( E\partial_p+\sum_{n=1}^{n_{max}}\mu_n\,\varphi_n(p) \Big) f(p) =
M(p) \sum_{n=1}^{n_{max}}a_n\,\varphi_n(p)
\end{equation}
where we have introduced the abbreviations
\begin{equation}
\mu_n:=\lambda_n\int\limits_{-\infty}^\infty\!dp'\,M(p')\,\varphi_n(p')
\quad\mbox{ and }\quad
a_n:=\lambda_n\int\limits_{-\infty}^\infty \!dp'\,f(p')\,\varphi_n(p')
\end{equation}
From its definition (3) we see that the collision frequency in this model
is given by $\nu(p)=\sum\mu_n\,\varphi_n(p)$. The solution of (9) 
vanishing at $p=\pm\infty$ is
\begin{equation}
f(p) = \frac{1}{E}\;e^{-N(p)/E}
\sum_{n=1}^{n_{max}}a_n\int\limits_{-\infty}^p\!
dp'\,\varphi_n(p')\,M(p')\,e^{N(p')/E}
\end{equation}
$N(p)$ (read ``capital $\nu$'') denotes the integral over the collision frequency
\begin{equation}
where N(p):=\int_0^p dq\,\nu(q)=\int_0^p dq\sum_{n=1}^{n_{max}}\mu_n\,\varphi_n(q)
\end{equation}
The coefficients $a_n$ must be determined selfconsistently by insertion of the
solution (11) into the definition of the coefficients (10). This leads to
the system of linear equations
\begin{equation}
\sum_{n=1}^{n_{max}}\Big( \mbox{A}_{mn}-\delta_{mn} \Big)\; a_n = 0
\end{equation}
where $\delta_{mn}$ denotes Kronecker's symbol and the matrix $\mbox{A}_{mn}$ 
is given by
\begin{displaymath}
\mbox{A}_{mn}:=\frac{1}{E}
\int\limits_{-\infty}^\infty\!dp\,e^{-N(p)/E}\lambda_m\,\varphi_m(p)
\int\limits_{-\infty}^p dp'\,\varphi_n(p')\,M(p')\,e^{N(p')/E}
\end{displaymath}
Partial integration in the $p$-integral using $N'(p)=\sum \mu_k\varphi_k(p)$ 
reveals that
\begin{displaymath}
\sum_{k=1}^{n_{max}}\frac{\mu_k}{\lambda_k}\Big( \mbox{A}_{kn}
-\delta_{kn} \Big) = 0
\end{displaymath}
so that the determinant of the matrix $(\mbox{A}-1)$ vanishes. Hence the 
homogeneous system (13) indeed has a nontrivial solution and the stationary
homogeneous Boltzmann equation is solved by (11).

\subsection{Asymptotic evaluation of the moments}
After changing the order of integrations we obtain for the $n$-th moment of $f$
given by (11) (remember that the solutions $a_k$ of (13) depend on $E$)
\begin{eqnarray}
\langle p^n \rangle (E) & = & \frac{\int_{-\infty}^\infty dp\,p^n f}
{\int_{-\infty}^\infty dp'\,f} =
\frac{\sum_{k=1}^{n_{max}}a_k(E)\,I_k^n(E)}
{\sum_{k=1}^{n_{max}}a_k(E)\,I_k^0(E)} \quad\quad\mbox{ with} \\[3mm]
I_k^n(E) & := & \int\limits_{-\infty}^\infty\!dp\,e^{-p^2\!/2\theta}
\varphi_k(p)\,e^{N(p)/E}\int\limits_p^\infty\!dq\,q^n\,e^{-N(q)/E} \nonumber
\end{eqnarray}
If we make the substitutions $t=N(q)$ and $s=N(p)$, we can interpret the 
double integrals in (14) as Laplace transforms with the Laplace variable 
$\varepsilon:=1/E$ (we denote the inverse function of $N$ by $N^{-1}$):
\mathindent1cm
\begin{eqnarray}
I_k^n(E) & = & \int\limits_{-\infty}^\infty\!ds\,
 \frac{\varphi_k\left(N^{-1}(s)\right)}{\nu\left(N^{-1}(s)\right)}\;
 e^{-\left(N^{-1}(s)\right)^2/2\theta}
 \int\limits_s^\infty\!dt\,
 \frac{\left(N^{-1}(t)\right)^n}{\nu\left(N^{-1}(t)\right)}\;
 e^{-(t-s)\varepsilon} \nonumber \\
& = & \int\limits_0^\infty\!dt\,e^{-t\varepsilon}
 \int\limits_{-\infty}^\infty\!ds\,
 \frac{\varphi_k\left(N^{-1}(s)\right)}{\nu\left(N^{-1}(s)\right)}\;
 \frac{\left(N^{-1}(t+s)\right)^n}{\nu\left(N^{-1}(t+s)\right)}\;
 e^{-\left(N^{-1}(s)\right)^2/2\theta}
\end{eqnarray}
\mathindent2cm
According to the Abelian theorems on Laplace transforms \cite{Doetsch}, the 
asymptotic behaviour of the Laplace transform (15) for small $\varepsilon$ is 
determined by the asymptotic behaviour of the inner integral for large $t$. 
Under the assumption that $N^{-1}(t)$ behaves asymptotically as given in 
(16) (see theorem 1 below), the
asymptotic behaviour of the $s$-integral for large $t$ can be determined in 
complete analogy to section 4 of reference \cite{Dalitz_moments}. The result is
\mathindent1cm
\begin{displaymath}
s\mbox{-integral in (15)}  \stackrel{t\to\infty}{\sim} 
 \frac{\left(N^{-1}(t)\right)^n}{\nu\left(N^{-1}(t)\right)}
 \underbrace{ \int_{-\infty}^\infty\!ds\, 
 \frac{\varphi_k\left(N^{-1}(s)\right)}{\nu\left(N^{-1}(s)\right)}\;
 e^{-\left(N^{-1}(s)\right)^2/2\theta} }_{\textstyle = \mu_k/\lambda_k}
\end{displaymath}
\mathindent2cm
For small values of $\varepsilon=1/E$, we may replace the $s$-integral in (15) by
this asymptotic expression for large $t$. If we resubstitute $t=N(p)$ and insert 
the result into (14), we obtain
\begin{displaymath}
\langle p^n\rangle (E)  \stackrel{E\to\infty}{\sim} 
\frac{\sum_{k=1}^{n_{max}}a_k(E)\,\frac{\mu_k}{\lambda_k}
      \int\limits_0^\infty\!dp\,p^n e^{-N(p)/E}}
     {\sum_{k=1}^{n_{max}}a_k(E)\,\frac{\mu_k}{\lambda_k}
      \int\limits_0^\infty\!dp\,e^{-N(p)/E}} =
\frac{\int\limits_0^\infty\!dp\,p^n e^{-N(p)/E}}
     {\int\limits_0^\infty\!dp\,e^{-N(p)/E}}
\end{displaymath}
This is exactly the same expression that was obtained in \cite{Dalitz_moments}
for kernels of rank one in the zero temperature limit. Consequently we
obtain the same high field behaviour of the moments (for details of the 
derivation see \cite{Dalitz_moments}):\\[5mm]
\begin{it}
{\bf Theorem 1:} The collision kernel be of the form (8) and the inverse
function $N^{-1}$ of the integral over the collision frequency $N(p)=\int_0^p
\nu(q)\,dq$ have the asymptotic behaviour
\begin{equation}
N^{-1}(t) \sim A\,t^\alpha L(t) \quad\mbox{ for }\quad t\to\infty
\end{equation}
where $\alpha>-1$ and $L$ be a continuous function with the property 
$\lim_{t\to\infty}L(\tau t)/L(t)=1$ for every $\tau>0$. \par
Then the $n$-th moment of the solution of the stationary homogeneous
Boltzmann equation behaves for $E\to\infty$ like
\begin{equation}
\langle p^n\rangle  \sim  \frac{A^n}{n+1}\frac{\Gamma(n\alpha+\alpha+1)}
{\Gamma(\alpha+1)}E^{n\alpha} L^n(E)
  \sim \mbox{const}\cdot\left(N^{-1}(E)\right)^n
\end{equation}
\end{it}
\vspace{2mm}
In particular the first moment or the drift velocity behaves as $N^{-1}(E)$ 
for large $E$. Because of $N(\infty)=\infty$, which is a necessary condition
for the existence of a stationary homogeneous solution of the Boltzmann
equation \cite{Poupaud}, we conclude that {\it velocity saturation is 
excluded for Hilbert Schmidt kernels of finite rank}.

\section{A conjecture}
Since we have seen that theorem 1 on the high field behaviour of the moments
is valid for Hilbert-Schmidt kernels of any finite rank, it is reasonable to
assume that our theorem holds for arbitrary Hilbert-Schmidt kernels as well.
This conjecture however is very difficult to prove; a major problem is that
the eigenfunctions are not positive for a rank greater than one, so that 
their sum (5) is not necessarily positive if it is truncated at finite $n$.
Maybe a proof is possible, if the kernel $K(p,p')$ is a continuous function
of $p$ and $p'$, for then the sum (5) converges even uniformly. Nevertheless
the problem mentioned above still remains.\par
As a reinforcement of our conjecture, let us consider the kernel of infinite
rank $K(p,p')=\rho |p-p'|$, for which  {\sc Gervois} and {\sc Piasecki} determined 
the large field behaviour of the first moment in \cite{Gervois}. They obtained
\begin{equation}
\langle p\rangle (E)\sim\sqrt{2E/\rho\pi}\quad\mbox{ for }E\to\infty
\end{equation}
which is independent of the temperature $\theta$. In this model, the
collision frequency reads 
\begin{displaymath}
\nu(p)=2\rho\theta\,M(p)+\rho |p|\left( 1-2\int_{|p|}^\infty M(p')\,dp'\right)
\end{displaymath}
Hence, the integral over the collision frequency behaves as $N(p)\sim\rho p^2\!/2$ 
for large $p$ and its inverse function as $N^{-1}(t)\sim\sqrt{2t/\rho}$ for large
$t$. Naive application of theorem 1 leads to
\begin{displaymath}
\langle p\rangle (E)\stackrel{E\to\infty}{\sim} \sqrt{2/\rho}\;
\frac{\Gamma(1+1)}{2\Gamma(0.5+1)}\sqrt{E}=\sqrt{2E/\rho\pi}
\end{displaymath}
which is identical with {\sc Gervois'} and {\sc Piasecki'}s result (18)!

\section{Optical phonon scattering}
Now let us consider collision models that take into account the quantum
nature of the lattice vibrations. Under the assumption of a single
phonon energy $\omega$ (this assumption holds for the optical mode), the 
general form of the collision kernel reads \cite{Niclot}
\begin{eqnarray}
M(p)\,K(p',p) & = & \beta(p',p)\Big\{ (N_o+1)\,\delta(p^2\!/2-p'^2\!/2+\omega) \\
    & & +\;N_o\,\delta(p^2\!/2-p'^2\!/2-\omega) \Big\} \nonumber
\end{eqnarray}
where $N_o=(e^{\omega/\theta}-1)^{-1}$ is the occupation number of the 
optical phonons with energy $\omega$ and the scattering cross
section $\beta(p',p)$ is a symmetric and positive function. The first
term in the curly braces describes phonon emission, the second phonon
absorption during the scattering process; the Dirac $\delta$-distributions
assure energy conservation.\par
Under the isotropy assumption $\beta=\beta(|p'|,|p|)$, we shall see that
an asymptotic solution of the Boltzmann equation is possible for high 
energy electrons with $p^2\!\gg\!2\omega$, provided the electric field is
sufficiently high $E\!\gg\!\omega$. The asymptotic knowledge of 
the electron distribution will be sufficient for the 
determination of the high field behaviour of the moments. \par
Our replacement of difference operators by
differential operators, which is the essential step in our procedure, is not
new. The first who did so were {\sc Yamashita} and {\sc Watanabe} 
\cite{Yamashita}. However, they obtained erroneous results for the electron
distribution and the drift velocity, which were corrected by {\sc Reik} and
{\sc Risken} \cite{Reik}. {\sc H\"ansch} and {\sc Schwerin} have applied
this method to inhomogeneous situations as well \cite{Haensch}. In all three
references, a combination of acoustic and optical phonon scattering is
considered and the scattering cross section $\beta$ is taken to be constant.

\subsection{The stationary homogeneous Boltzmann equation}
Under the isotropy assumption $\beta(p',p)=\beta(|p'|,|p|)$, the one dimensional
stationary homogeneous Boltzmann equation with the collision kernel (19) reads
\begin{eqnarray}
\lefteqn{\Big\{ E\partial_p + (N_o+1)\Theta(p^2-2\omega)\beta^-(p)
+ N_o\beta^+(p) \Big\} f(p)} \nonumber \\[2mm]
 & & = (N_o+1)\,\beta^+(p)\,\frac{1}{2} \left\{ f\left(-\sqrt{p^2+2\omega}\right)
       +f\left(+\sqrt{p^2+2\omega}\right) \right\} \\[2mm]
 & & +\;N_o\Theta(p^2-2\omega)\,\beta^-(p)\,\frac{1}{2}
     \left\{ f\left(-\sqrt{p^2-2\omega}\right)
     +f\left(+\sqrt{p^2-2\omega}\right) \right\} \nonumber
\end{eqnarray}
where $\Theta$ denotes Heaviside's step function, and the abbreviation 
$\beta^\pm$ means
\begin{displaymath}
\beta^\pm(p) := 2\beta\left( \sqrt{p^2\pm 2\omega}, p\right) \Big/ 
                \sqrt{p^2\pm 2\omega}
\end{displaymath}
Now let us split $f(p)=f_o(p)+f_e(p)$ into its odd and even part 
$f_o(p):=(f(p)-f(-p))/2$ and $f_e(p):=(f(p)+f(-p))/2$. Then we may write 
(20) as two coupled equations for $f_o$ and $f_e$ for $p>0$. With the
substitution $q:=p^2\!/2$ and the notations
\begin{equation}
F_{\stackrel{\scriptstyle o}{e}} (q):=f_{\stackrel{\scriptstyle o}{e}} 
\left(\sqrt{2q}\right) = f_{\stackrel{\scriptstyle o}{e}} (p)=
\frac{1}{2}\Big( f(p)\mp f(-p)\Big) \quad\quad\mbox{ and}
\end{equation}
\begin{equation}
B(q):= \beta\left(\sqrt{2(q-\omega)}, \sqrt{2q}\right) \Big/
       \sqrt{q(q-\omega)}
\end{equation}
we obtain the coupled system of equations for $F_{\stackrel{\scriptstyle o}{e}}$
(remember that $\beta^\pm$ are even functions of $p$ and that $\beta(p',p)$ is 
symmetric with respect to exchange of $p$ and $p'$)
\begin{subeqnarray}
E\partial_q F_o(q) & = & (N_o+1) B(q+\omega) F_e(q+\omega) + 
                      N_o\Theta(q-\omega)B(q) F_e(q-\omega) \\
 & & -\;\Big[ (N_o+1)\Theta(q-\omega)B(q) + N_o B(q+\omega)\Big] F_e(q)
   \nonumber \\[2mm]
E\partial_q F_e(q) & = & -\Big[(N_o+1)\Theta(q-\omega)B(q)+N_oB(q+\omega)
                          \Big] F_o(q)  
\end{subeqnarray}
Using (21), we may express the moments of the electron distribution in terms 
of the functions $F_{\stackrel{\scriptstyle o}{e}}$
\begin{equation}
\langle p^n \rangle (E)=\frac{\int_{-\infty}^\infty dp\,p^n f}
{\int_{-\infty}^\infty dp\,f} = \left\{
\begin{array}{ll}
\frac{\int_0^\infty dq\,(2q)^{\frac{n-1}{2}}F_o(q)}
{\int_0^\infty dq\,F_e(q)/\sqrt{2q}} & \mbox{ for odd }n \\[4mm]
\frac{\int_0^\infty dq\,(2q)^{\frac{n-1}{2}}F_e(q)}
{\int_0^\infty dq\,F_e(q)/\sqrt{2q}} & \mbox{ for even }n 
\end{array}\right.
\end{equation}

\subsection{Asymptotic solution for fast electrons}
For sufficiently high electric fields, an asymptotic solution of (23)
in the region $p^2\!\gg\!2\omega$ is possible via replacement of the
difference operators by differential operators. For this purpose we switch
to the new varible $x\!:=\!q/E$ and denote the resulting new functions with
hats:
\begin{displaymath}
\hat{F}_{\stackrel{\scriptstyle o}{e}} (x):=F_{\stackrel{\scriptstyle o}{e}} (Ex)
=F_{\stackrel{\scriptstyle o}{e}} (q) \quad\mbox{ and }\quad
\hat{B}(x):=B(Ex)=B(q)
\end{displaymath}
After a rearrangement of terms, the system (23) reads for 
$x>\epsilon:=\omega/E$ 
\begin{subeqnarray}
\partial_x\hat{F}_o(x) & = & (N_o+1)\Big\{ \hat{B}(x+\epsilon) \hat{F}_e
                             (x+\epsilon)-\hat{B}(x) \hat{F}_e(x) \Big\} \\
             & & -\;N_o \Big\{ \hat{B}(x+\epsilon)\hat{F}_e(x)-
                      \hat{B}(x)\hat{F}_e(x-\epsilon) \Big\} \nonumber \\[2mm]
\partial_x\hat{F}_e(x) & = & -\;\Big\{ (N_o+1)\hat{B}(x)+N_o\hat{B}(x+\epsilon) 
                              \Big\} \hat{F}_o(x)
\end{subeqnarray}
For small $\epsilon\ll 1$ and large $x\gg\epsilon$ (which means $p^2\!\gg 2\omega$),
the difference operators on the right hand side can be replaced by differential 
operators. If we make a Taylor expansion of $\hat{F}_e(x\pm\epsilon)$ and 
$\hat{B}(x+\epsilon)$ around
$x$ and keep only terms up to first order in $\epsilon$, we obtain
\begin{subeqnarray}
\partial_x\hat{F}_o(x) & \approx & 
                 \epsilon\partial_x\Big(\hat{B}(x)\hat{F}_e(x)\Big) \\
\partial_x\hat{F}_e(x) & \approx & -\Big\{ (2N_o+1)\hat{B}(x)
                         +\epsilon N_o\partial_x\hat{B}(x)\Big\}\hat{F}_o(x)
\end{subeqnarray}
Taking into account that $\hat{F}_{\stackrel{\scriptstyle o}{e}}(\infty)=0$,
we may integrate (26a) from $x$ to infinity and insert the result into 
(26b). Again we keep only terms up to first order in $\epsilon$.
\begin{subeqnarray}
\hat{F}_o(x) & \approx & \epsilon \hat{B}(x)\hat{F}_e(x) \\
\partial_x\hat{F}_e(x) & \approx & -\epsilon (2N_o+1)
                         \hat{B}^2(x)\hat{F}_e(x)
\end{subeqnarray}
Equation (27b) easily can be integrated with the boundary condition 
$\hat{F}_e(\infty)=0$. Thus we obtain an asymptotic solution of (25)
for $x\gg\epsilon$ under the assumption that $\epsilon=\omega/E$ be 
small. We may also take into account the region where $x\gg\epsilon$ does
not hold by introduction of two functions $\varrho_{\stackrel{\scriptstyle o}{e}}
(t)$ with the property
$\lim_{t\to\infty}\varrho_{\stackrel{\scriptstyle o}{e}}(t)=1$. Then the
factors $\varrho_{\stackrel{\scriptstyle o}{e}}(x/\epsilon)$ describe 
transition to the asymptotic domain $x\gg\epsilon$ and the
solution globally valid in $x$ reads
\begin{subeqnarray}
\hat{F}_e(x) & \stackrel{\epsilon\ll 1}{\approx} & C\varrho_e(x/\epsilon)
\,\exp\left(\epsilon(2N_o+1)\int_0^x\!dy\,\hat{B}^2(y)\right) \\
\hat{F}_o(x) & \stackrel{\epsilon\ll 1}{\approx} & C\varrho_o(x/\epsilon)
\,\epsilon\hat{B}(x)\,\exp\left(\epsilon(2N_o+1)\int_0^x\!dy\,\hat{B}^2(y)\right) 
\end{subeqnarray}
with some arbitrary constant $C>0$.

\subsection{High field behaviour of the moments}
If we insert the high field solution (28) into the expression
for the moments (24), we find after the substitutions $q=Ex$ and 
$q'=Ey$ (remember that $\epsilon=\omega /E$)
\begin{equation}
\langle p^n \rangle (E) \stackrel{E\gg\omega}{\approx} I_n(E) / I_0(E)
\quad\quad\mbox{ with}
\end{equation}
\begin{equation}
I_n(E)=\left\{
\begin{array}{ll}
\frac{\omega}{E}\int\limits_0^\infty\!dq\,(2q)^{\frac{n-1}{2}}
\varrho_o\left(\frac{q}{\omega}\right)\,B(q)\,
\exp\left(-\frac{\omega (2N_o+1) A(q)}{E^2}\right)
 & \mbox{ for odd }n \\[2mm]
\int\limits_0^\infty\!dq\,(2q)^{\frac{n-1}{2}}
\varrho_e\left(\frac{q}{\omega}\right)\,
\exp\left(-\frac{\omega (2N_o+1) A(q)}{E^2}\right) 
 & \mbox{ for even }n
\end{array}\right.
\end{equation}
where we have introduced the abbreviation
\begin{equation}
A(q) := \int_0^q dq'\,B^2(q')
\end{equation}
As $E$ goes to infinity, the integrals in the numerator and denominator of (29)
diverge, whilst the integrals remain finite if the upper limit is replaced by 
a finite value. Hence their behaviour for large fields is determined by the 
asymptotic behaviour of the integrands for large $q$ and consequently our 
knowledge of the asymptotic form of the electron distribution for 
$p^2\!\gg 2\omega$ will suffice for the determination of the high field 
moments. \par
For convenience, let us assume that $A(\infty)=\infty$. Then we can procede as 
in section 3 by writing the integrals in (30) as Laplace transforms, which we
can evaluate with the Abelian theorem again. To this purpose we make the
substitution $t=A(q)$ and obtain for large values of $E$ (the inverse function 
of $A$ is denoted by $A^{-1}$)
\begin{equation}
I_n(E) \sim \left\{
\begin{array}{ll}
\frac{\textstyle \omega}{\textstyle E}\int\limits_0^\infty\!dt\,
\frac{\textstyle \Big(2A^{-1}(t)\Big)^\frac{n-1}{2}}
{\textstyle B\Big( A^{-1}(t)\Big)}
\;\exp\left(-\frac{\omega (2N_o+1)t}{E^2}\right) 
 & \mbox{ for odd }n \\[5mm]
\int\limits_0^\infty\!dt\,
\frac{\textstyle \Big(2A^{-1}(t)\Big)^\frac{n-1}{2}}
{\textstyle B^2\Big( A^{-1}(t)\Big)}
\;\exp\left(-\frac{\omega (2N_o+1)t}{E^2}\right) 
 & \mbox{ for even }n 
\end{array}\right.
\end{equation}
Under the assumption that $B(q)$ asymptotically grows like a power for 
large values of $q$,
the asymptotic large field behaviour of the integrals in (32) is easily 
evaluated with the aid of the Abelian theorem on Laplace transforms 
(\cite{Doetsch} part I, chapter 13 \S1 theorem 7). The elementary 
calculation leads to\\[5mm]
\begin{it}
{\bf Theorem 2:} The collision operator be of the form (19) and the 
scattering cross section $\beta$ satisfy the isotropy property
$\beta(p',p)=\beta(|p'|,|p|)$ and be such that for $q\to\infty$
\begin{equation}
B(q) = \beta\left(\sqrt{2(q-\omega)}, \sqrt{2q}\right) \Big/
       \sqrt{q(q-\omega)} \quad \sim \quad a\cdot q^\alpha 
\end{equation}
with some constants $a>0$ and $\alpha\ge -1/2$. \par
Then the $n$-th moment of the solution of the stationary homogeneous
Boltzmann equation behaves for $E\to\infty$ like
\begin{equation}
\langle p^n \rangle (E) \sim \left\{
\begin{array}{ll}
2^{\frac{n}{2}}a\,\omega\,
\frac{\textstyle \Gamma\left(\frac{n+1+2\alpha}{4\alpha+2}\right)}
{\textstyle \Gamma\left(\frac{1}{4\alpha+2}\right)}
\left(\frac{\textstyle 2\alpha+1}
{\textstyle a^2\omega\,(2N_o+1)}\right)^{\frac{n+2\alpha}{4\alpha+2}}
E^{\frac{n-1}{2\alpha+1}} & \mbox{for odd }n \\[7mm]
2^{\frac{n}{2}}\frac{\textstyle \Gamma\left(\frac{n+1}{4\alpha+2}\right)}
{\textstyle \Gamma\left(\frac{1}{4\alpha+2}\right)}
\left(\frac{\textstyle 2\alpha+1}
{\textstyle a^2\omega\,(2N_o+1)}\right)^{\frac{n}{4\alpha+2}}
E^{\frac{n}{2\alpha+1}} & \mbox{for even }n 
\end{array}\right.
\end{equation}
\end{it}
\vspace{2mm}
Particularly, for the first moment we find
\begin{equation}
\langle p\rangle (E)\sim \frac{\Gamma\left(\frac{\alpha+1}{2\alpha+1}\right)}
{\Gamma\left(\frac{1}{4\alpha+2}\right)}\,
\sqrt{\frac{(4\alpha+2)\omega}{2N_o+1}}
\end{equation}
which is a finite value. Consequently the {\it drift velocity saturates if the 
electrons interact with optical phonons!}

\subsection{An example}
For the particular choice $\beta(p,p')=C|p|\cdot|p'|$ proposed by {\sc Mahan}
\cite{Mahan}, the function $B(q)$ defined by (22) is simply a constant, so that
the stationary homogeneous Boltzmann equation can be solved exactly in 
the zero temperature limit $N_o=0$ \cite{Dalitz_exact}. The resulting drift 
velocity is
\begin{equation}
\langle p\rangle (E) = \frac{\omega/zE}
{\sqrt{2\omega}+e^{z\omega}\sqrt{\pi/2z}\:\mbox{erfc}\left(\sqrt{z\omega}\right)}
\end{equation}
where $z$ is the positive real root of the transcendental equation
$(zE)^2=1-e^{-z\omega}$. For large $E$, this solution is approximately 
$z(E)\approx\omega/E^2$ so that the current approaches the finite value 
$j_{sat}=\sqrt{2\omega/\pi}$ as $E$ goes to infinity. This result is in agreement
with (35) for $N_o=0$ and $\alpha=0$. \par
In this model the asymptotic formula for the drift velocity is easily improved
by keeping terms up to second order in (25). If we do so, we find for the first 
moment
\begin{equation}
\langle p\rangle(E) \sim \sqrt{\frac{2\omega}{\pi (1+\omega^2\!/2E^2)}}
\end{equation}
A comparison of our asymptotic formula (37) with the exact value (36) is 
given in the figure. The agreement for large values of the electric field 
$E$ is clearly visible. (The figure is omitted in the cond\_mat-preprint).

\section{Summary and conclusions}
In this article we have seen that the experimenatlly observed 
saturation of the electron drift velocity in semiconductors cannot
be obtained with regular collision kernels of finite rank. This
result seems to remain valid even for infinite rank, though we could
not present a rigorous proof of this conjecture. An extension of our 
argumentation
to three dimensions is straightforward and can be performed in complete
analogy to the case of rank one \cite{Dalitz_moments}. \par
In his investigation of the null space of the Boltzmann collision
operator \cite{Majorana}, {\sc Majorana} has already observed
that singular collision kernels have mathematical properties rather
different from regular collision kernels. In this article we have
seen that also velocity saturation is a feature restricted to singular
kernels. \par
Under the assumption of isotropic optical phonon scattering we were
able to derive velocity saturation, regardless of the detailed form
of the scattering cross section. However two interesting questions
remain, which are worth further investigation. First, is there
a generalization of our proof to three dimensions beyond the framework
of a two term Legendre approximation? The latter approximation, used
in \cite{Yamashita} and \cite{Reik}, is an effective one dimensional
approximation, which is equivalent to our one dimensional model.
Second, does our result remain unchanged if the scattering is anisotropic? 

\section*{Acknowledgement}
The author is grateful to the ``Studienstiftung des deutschen Volkes''
for financial support of this work.

\end{document}